\documentclass{llncs}
                
\usepackage{amsmath}
\usepackage{qtree}
\usepackage{multicol}

\begin{document}

\title{Language Approximation With One-Counter Automata}

\author{Alexander Sakharov}

\institute{}

\maketitle

\begin{abstract}
We present a method for approximating context-free languages with one-counter automata. This approximation allows the reconstruction of parse trees of the original grammar. We identify a decidable superset of regular languages whose elements, i.e. languages, are recognized by one-counter automata.  
\end{abstract}

\newenvironment{algfont}{\fontfamily{cmtt}\selectfont}{\par}

\section{Introduction}

Many applications utilize grammars that evolve, and it is not possible to assess grammar properties. Context-free (CF) parsing is difficult with ambiguous grammars. The time complexity of parsing CF languages is cubic in the size of input in a general case, i.e. when the properties of the grammar are unknown. This time complexity is prohibitive for some applications. In comparison to CF languages, regular languages can be parsed in linear time by finite automata. 

The approximation of CF languages is a problem that has been extensively studied because of its importance in a number of applications. CF languages are usually approximated with regular languages \cite{Mohri01}. Some advanced approximation methods have been developed \cite{Egecioglu09}. Approximating languages are usually supersets of the source CF languages. 

There are several problems associated with regular approximation of CF languages. First, this approximation is inaccurate. Second, and most importantly, parse trees are mainly unavailable. Parse trees embody the syntactic information about the input. The primary goal of parsing is to extract this syntactic information. Superset approximation does not assist in deciding whether an input belongs to the source language; it only helps establish that an input does not belong to the source language. 

It is claimed in \cite{Mohri01} that the approximating finite automata can be used for parsing, but the parse trees reconstructed from acceptance transition sequences of the approximating automaton essentially differ from the parse trees of the source language, and some acceptance transition sequences cannot be mapped to parse trees. Another method for reconstructing parse trees from accepting transition sequences of the approximating automata was proposed in \cite{Nederhof98}. The problem with this method is that the reconstruction requires cubic time, which defeats the purpose of language approximation.

One-counter (OC) languages are the languages recognized by OC automata. These languages are an important subclass of CF languages and a proper superset  of regular languages. OC languages are utilized in program verification \cite{Bouajjani11} and XML validation \cite{Chitic04}. Stochastic one-counter automata have applications in Markov decision processes \cite{Brazdil10} and stochastic games \cite{BBE10}. Since OC languages do not enjoy a characterization via grammars \cite{Autebert97}, OC automata have not been used much for parsing. 

This paper describes the approximation of CF languages with OC automata, which has not been much investigated before. OC automata provide a more accurate approximation of CF grammars than finite automata. The approximating languages are supersets the source CF languages. We show how to generate parse trees of the source grammar from acceptance transition sequences of the recognizing OC automaton. Our approximation method is well suited for ambiguous languages. This method can also be used for the approximation with finite automata but the ability to reconstruct parse trees is lost in the latter case.

OC language parsing is done in quadratic time of the size of input \cite{Greibach75}. Parse trees of the source grammar can be reconstructed from acceptance transition sequences of the approximating OC automaton in linear time.

The approximation with OC automata is useful even for the inputs that belong to the approximating language but not to the source language because the parse trees are available for such inputs. Some of these inputs are just erroneous inputs. Having parse trees for these erroneous inputs may be quite useful. It is somewhat similar to error recovery in parsing \cite{Aho06}.

We identify a decidable set of grammars whose languages are recognized by OC automata. These languages constitute a proper superset of regular languages. They significantly overlap with input-driven languages \cite{Okhotin14}.

\section{Preliminaries} 

OC automata are defined as an extension of finite automata. Their transitions have the form:

\noindent
$s, t, c \rightarrow r, a$

\noindent 
where $s$ and $r$ are states, $t$ is a terminal, $c \in \{ 0, + \}$, $a \in \{+1,-1, 0 \}$. Transitions with $c = 0$ apply when the counter is zero, and transitions with $c = +$ apply when the counter is nonzero. The value of $a$ defines how the counter value changes. Transitions with $c = 0$ and $a = -1$ are disallowed. It is usually required for the counter to be zero at a final state in order for the input to be accepted. Alternatively, OC automata can be defined as pushdown automata with a single stack symbol. We consider non-deterministic automata.

OC automata can be directly used for parsing, i.e. without engaging grammars. Trees can be generated from acceptance transition sequences. It is fair to call them parse trees because these trees embody structural information about the input. Terminals and source states of transitions decreasing the counter serve as labels for leaf nodes of these trees. Source states of transitions retaining or increasing the counter serve as labels for non-leaf nodes. This interpretation is based on the intuitive assumption that any counter-increasing transition opens a construct, and a matching counter-decreasing transition closes it. Transitions retaining the counter value are treated as right-linear constructs.

\section{Lax Input-Driven Languages}

Consider languages defined by productions of the following three forms:

\noindent
1. $A \rightarrow t B$ $\quad \quad \quad \quad$ 2. $A \rightarrow u B v C$ $\quad \quad \quad \quad$ 3. $A \rightarrow \epsilon$

\noindent
Nonterminals having production $A \rightarrow \epsilon$, where $\epsilon$ denotes the empty string, are called nullable. These languages are a superset of input-driven languages \cite{Okhotin14}. We call them lax input-driven (LID) languages. Input-driven languages are defined by productions of these three forms and are subject to one additional condition: the set of terminals $T$ of any input-driven language is the union of three disjoint sets $T_0$, $T_+$, and $T_-$. Terminals $t$ from productions $A \rightarrow t B$ should belong to $T_0$. Terminals $u$ from productions $A \rightarrow u B v C$ should belong to $T_+$, and $v$ should belong to $T_-$.

Input-driven languages themselves are a wide class of languages. For instance, they include the languages of balanced grammars \cite{Berstel02}. It is reasonable to abandon the assumption about the three disjoint sets of terminals for stochastic languages because set membership should not be deterministic in a probabilistic setting. LID languages can be approximated by OC automata. We specify a subset of LID languages that are recognized by OC automata. This subset is a proper superset of regular languages. 

We build an automaton with state $A$ for every nonterminal $A$. Let $\mathcal{R}(A)$ denote the set consisting of $A$ and all such $B$ that $A =>^* ... B$  is a valid derivation. A simple iterative procedure can determine the existence of these derivations for all nonterminal pairs. Every production $A \rightarrow u B$ maps to transitions $A, u, 0 \rightarrow B, 0$ and $A, u, + \rightarrow B, 0$. Every production $A \rightarrow u B v C$ maps to transitions $A, u, 0 \rightarrow B, +1$ and $A, u, + \rightarrow B, +1$. Additionally, transitions $D, v, + \rightarrow C, -1$ are created for all nullable nonterminals $D \in \mathcal{R}(B)$. Start nonterminal $S$ is the start state. State $E$ is final iff it is nullable and $E \in \mathcal{R}(S)$. Let $\mathcal{L}$ denote the language defined by a grammar or automaton.  

In order to build a parse tree, we iterate over transitions in an acceptance sequence of an approximating OC automaton and maintain the stack of nonterminals. For any transition $A, u, c \rightarrow B, 0$, $u$ and $B$ are the children of A.  For any transition $A, u, c \rightarrow B, +1$, $u$ and $B$ are the first two children of $A$. $A$ is pushed onto the stack. For any transition $C, v, + \rightarrow D, -1$, a nonterminal is popped from the stack, $v$ and $D$ are added as additional children of that node. 

\textbf{Proposition 1.} \textit{If automaton $\Omega$ is built from LID grammar $\Gamma$ by the above rules, then $\mathcal{L}(\Gamma) \subseteq \mathcal{L}(\Omega)$.}

Consider the parse tree of an input string from $\mathcal{L}(\Gamma)$. Let us traverse this parse tree in pre-order. Every node is visited once during this traversal. Note that nonterminals and terminals alternate in the traversal sequence. Every triple $A,b,C$ in the sequence (where $b$ is a terminal) corresponds to a transition generated from the grammar. The counter is zero at the end. Therefore, the input is accepted by $\Omega$. \qed

\textbf{Proposition 2.} \textit{Suppose automaton $\Omega$ is built from LID grammar $\Gamma$. If for any two productions $A \rightarrow b B c C$ and $E \rightarrow f F g G$, either $c = g$ and $C = G$, or $\mathcal{R}(B) \cap \mathcal{R}(F) = \emptyset$, then $\mathcal{L}(\Gamma) = \mathcal{L}(\Omega)$.}

Let $\triangleright(t)$ and $\triangleleft(t)$ denote the source and destination state (nonterminal) of transition $t$, respectively. We prove by induction on the number of counter-increasing transitions that if $t_1,...,t_n$ is a counter-balanced sequence of transitions of $\Omega$ for input string $s_1,...,s_n$, and $\triangleleft(t_n)$ is nullable, then $\triangleright(t_1) \Rightarrow^* s_1,...,s_n$ is a valid derivation in $\Gamma$, and $\triangleleft(t_n) \in \mathcal{R}(\triangleright(t_1))$. Proposition 2 is a straightforward corollary of this.

Base: Clearly, this proposition holds for sequences without counter-increasing transitions.
 
Induction step: Suppose the proposition holds for sequences with not more than $m$ counter-increasing transitions. Consider the first counter-increasing transition $t_i$. Let $t_j$ be its balancing counter-decreasing transition. 

Suppose $t_i$ and $t_j$ correspond to productions $A \rightarrow s_i B c C$ and  $E \rightarrow f F s_j G$, respectively. By the induction assumption, transition sequence $t_{i+1},...,t_{j-1}$ maps to derivation $B \Rightarrow^* s_{i+1},...,s_{j-1}$, and $\triangleleft(t_{j-1}) \in \mathcal{R}(B)$. Note that $\triangleleft(t_{j-1}) \in \mathcal{R}(F)$ as well. Therefore, $c = s_j$, $C = G$, and $t_j$ is identical to the counter-decreasing transition generated from production $A \rightarrow s_i B c C$. By the induction assumption, $C \Rightarrow^* s_{j+1},...,s_n$ is a valid derivation, and $\triangleleft(t_n) \in \mathcal{R}(C)$. Hence, $A \Rightarrow^* s_i,...,s_n$. Transitions $s_1,...,s_{i-1}$ are all counter-retaining, and thus, $A \in \mathcal{R}(\triangleright(t_1))$.  Therefore, $\triangleright(t_1) \Rightarrow^* s_1,...,s_n$, and $\triangleleft(t_n) \in \mathcal{R}(\triangleright(t_1))$. \qed
 
The automata generated from LID languages may accept input strings outside of the language defined by that grammar. The trees reconstructed from acceptance transition sequences of the approximating OC automaton may not match productions of the source LID grammar. Nonetheless, the proof of Proposition 2 shows that the trees reconstructed from acceptance transition sequences are the parse trees of the source grammar for the grammars satisfying the condition of Proposition 2. The set of languages defined by the grammars satisfying the condition of Proposition 2 is a proper superset of regular languages because it contains language $\{ a^n b^n : n \geq 0 \}$. 

If a LID grammar satisfies the condition of Proposition 2, then there is no more than one distinct production $A \rightarrow b B c C$ for any triple $A,b,B$. For any nullable nonterminal $D$ from a LID grammar satisfying the condition of Proposition 2, there is no more than one such pair $c,C$ that there is production $A \rightarrow b B c C$ where $D \in \mathcal{R}(B)$. We call $c,C$ the ancestor pair of $D$. 

\textbf{Definition.} \textit{Nonterminal $A$ from a LID grammar is called regular if no $D \in \mathcal{R}(A)$ has productions of form 2, or if all nonterminals from the right-hand side of every $A$ production are regular.}

This definition reflects the fact that the part of the grammar related to nonterminal $A$ is regular. The productions of this part are comprised of the $A$ productions and the productions of $A$ descendants. $A$ is the start nonterminal of this sub-grammar.

\textbf{Proposition 3.} \textit{If for any two productions $A \rightarrow b B c C$ and $E \rightarrow f F g G$ of LID grammar $\Gamma$ where $c \neq g$ or $C \neq G$, every nonterminal $D \in \mathcal{R}(B) \cap \mathcal{R}(F)$ is regular, then there exist a OC automaton $\Omega$ such that $\mathcal{L}(\Gamma) = \mathcal{L}(\Omega)$.}

Let us eliminate such productions $A \rightarrow b B c C$ that no $D \in \mathcal{R}(B)$ has productions of form 2. First, we replicate all $D \in \mathcal{R}(B)$ along with their productions. Then we replace productions $E \rightarrow \epsilon$ with productions $E \rightarrow c C$ for all replicated nonterminals $E$. Finally, we replace $A \rightarrow b B c C$ with $A \rightarrow b B'$ where $B'$ is the replica of $B$. Clearly, this transformation does not change the language defined by the grammar. This transformation does not affect any other production of form 2, and no new productions of form 2 are created. By applying this transformation iteratively, we replace  all productions $A \rightarrow b B c C$ where $A$ is regular with productions of forms 1 and 3. 

Note that this transformation may be limited to the nonterminals from $\mathcal{R}(B) \cap \mathcal{R}(F)$ for production pairs $A \rightarrow b B c C$, $E \rightarrow f F g G$. As a result, we get LID grammar $\Gamma'$ defining the same language and in which no nonterminal $D \in \mathcal{R}(B) \cap \mathcal{R}(F)$ for any production pair $A \rightarrow b B c C$, $E \rightarrow f F g G$ has productions of form 2.

For every production $A \rightarrow b B c C$ from $\Gamma'$, consider the set of all $D \in \mathcal{R}(B)$ that also belong to $\mathcal{R}(F)$ for at least one other production $E \rightarrow f F g G$ such that $c \neq g$ or $C \neq G$. We replicate these $D$ altogether along with their productions. After that, we replace $B$ in production $A \rightarrow b B c C$ by its replica. Note that no production of form 2 is affected by this replication, neither new productions of form 2 are created. Now $\mathcal{R}(B)$ does not intersect with $\mathcal{R}(F)$. Clearly, this transformation does not change the language defined by the source grammar. 

We repeat this transformation for the remaining productions of form 2 in $\Gamma'$. Note that the $\mathcal{R}$ set for any of these productions of form 2 does not intersect with the respective set of any production for which this transformation has been completed before. Grammar $\Gamma''$ obtained as the result of this transformation satisfies the condition of Proposition 2, and the automaton generated from the transformed grammar recognizes the language defined by this grammar. \qed

Proposition 3 gives a sufficient condition for a LID language to be recognizable by a OC automaton. This condition is decidable and can be efficiently verified. This proof of Proposition 3 basically gives an algorithm for building a grammar whose OC automaton recognizes the language of the source LID grammar. The states of this automaton can be traced back to the nonterminals of the source grammar. However, the transformation from the proof of Proposition 3 will likely lead to an explosion in the number of states.

\section{Approximation of Context-Free Languages}

The same technique could be used to approximate all CF languages by OC automata. Without loss of generality, we can assume that CF grammar productions are in the Greibach normal form (GNF), i.e. every production is $A \rightarrow b B_1 ... B_k$ where $k \ge 0$. Again, we consider automata whose states are grammar nonterminals. Start nonterminal $S$ is the start state. There is one and only final state $Z$ that does not map to any nonterminal. Transitions are constructed as follows:

\noindent
For every production $A \rightarrow b B_1$:

$\quad \quad A, b, 0 \rightarrow B_1, 0$ $\quad \quad \quad \quad A, b, + \rightarrow B_1, 0$

\noindent
For every production $A \rightarrow b B_1 ... B_k$ where $k > 1$:

$\quad \quad A, b, 0 \rightarrow B_1, +1$ $\quad \quad \quad \quad A, b, + \rightarrow B_1, +1$

\noindent
For every production $A \rightarrow b B_1 ... B_k$, $n \in \{1,...,k-1\}$, and production $D \rightarrow d$ such that $D \in \mathcal{R}(B_{n-1})$:

$\quad \quad D, d, + \rightarrow B_n, 0$

\noindent
For every production $A \rightarrow b B_1 ... B_k$ and production $D \rightarrow d$ such that $D \in \mathcal{R}(B_{k-1})$:

$\quad \quad D, d, + \rightarrow B_k, -1$

\noindent
For every production $D \rightarrow d$ such that $D \in \mathcal{R}(S)$:

$\quad \quad D, d, 0 \rightarrow Z, 0$

\textbf{Proposition 4.} \textit{If automaton $\Omega$ is built from CF grammar $\Gamma$ by the above rules, then $\mathcal{L}(\Gamma) \subseteq \mathcal{L}(\Omega)$.}

The proof of this proposition is similar to the proof of Proposition 1. We traverse the parse tree of an input string from $\mathcal{L}(\Gamma)$ in pre-order. Again, nonterminals and terminals alternate in the traversal sequence. Every triple $A,b,C$ in the sequence corresponds to a transition generated from the grammar, and the counter is zero at the end of the traversal.

The reconstruction of trees from acceptance transition sequences is done slightly different for CF grammars. We mark transitions $D, d, + \rightarrow B_n, 0$ in order to distinguish them from transitions $A, b, + \rightarrow B_1, 0$. For any unmarked transition $A, u, c \rightarrow B, 0$, $u$ and $B$ are the children of A.  For any marked transition $D, d, + \rightarrow B_n, 0$ generated from production $A \rightarrow b B_1 ... B_k$ where $n \leq k-1$, $B_n$ is added as an additional child to the top nonterminal on the stack. For any transition $A, u, c \rightarrow B, +1$, $u$ and $B$ are the first two children of $A$, and $A$ is pushed onto the stack. For any transition $C, v, n \rightarrow D, -1$, $v$ becomes the only child of $C$, a nonterminal is popped from the stack, $D$ is added as an additional child of that node. 

As an example, consider a grammar of arithmetic expressions in GNF:
 
\noindent
$E \rightarrow i$ $\quad \quad \quad$ $E \rightarrow i P$ $\quad \quad \quad$ $E \rightarrow ( E R$ $\quad \quad \quad$ $E \rightarrow ( E R P$
 
\noindent
$P \rightarrow + E$ $\quad \quad \quad$ $P \rightarrow * T$ $\quad \quad \quad$ $P \rightarrow * T L E$ 
 
\noindent
$T \rightarrow i$ $\quad \quad \quad$ $T \rightarrow i Q$ $\quad \quad \quad$ $T \rightarrow ( E R$ $\quad \quad \quad$ $T \rightarrow ( E R Q$

\noindent
$Q \rightarrow * T$ $\quad \quad \quad$ $R \rightarrow )$ $\quad \quad \quad$ $L \rightarrow +$

Note that this grammar can be easily turned into a LID grammar. Here is the automaton generated from the above grammar where $n = 0, +$:

\noindent
$E, i, n \rightarrow P, 0$ $\quad \quad$ \underline{$E, i, + \rightarrow R, 0$}  
$\quad \quad$
$E, i, + \rightarrow R, -1$ $\quad \quad$ $E, i, 0 \rightarrow Z, 0$ 

$E, (, n \rightarrow E, +1$ 
 
\noindent
$T, i, + \rightarrow R, -1$ $\quad \quad$ $T, i, + \rightarrow R, 0$ $\quad \quad$ \underline{$T, i, + \rightarrow L, 0$}
$\quad \quad$ $T, i, n \rightarrow Q, 0$ 

$T, i, 0 \rightarrow Z, 0$ $\quad \quad$ $T, (, n \rightarrow E, +1$ 

\noindent
$R, ), + \rightarrow R, -1$ $\quad \quad$ $R, ), + \rightarrow R, 0$ 
$\quad \quad$
$R, ), + \rightarrow P, -1$ $\quad \quad$ $R, ), + \rightarrow Q, -1$ 

$R, ), + \rightarrow L, 0$ $\quad \quad$ $R, ), 0 \rightarrow Z, 0$

\noindent
$P, +, n \rightarrow E, 0$ $\quad \quad$ $P, *, n \rightarrow T, 0$ $\quad \quad$ $P, *, n \rightarrow T, +1$

\noindent
$Q, *, n \rightarrow T, 0$ $\quad \quad$ $L, +, + \rightarrow E, -1$

Marked transitions are underlined. Consider input $a * d + (b + c)$. An acceptance transition sequence for this input along with the tree reconstructed from the sequence are shown below:

\begin{multicols}{2}
$E,a,0 \rightarrow P,0$
 
$P,*,0 \rightarrow T,+1$
 
\underline{$T,d,+ \rightarrow L,0$}

$L,+,+ \rightarrow E,-1$
 
$E,(,0 \rightarrow E,+1$
 
$E,b,+ \rightarrow P,0$
 
$P,+,+ \rightarrow E,0$
 
$E,c,+ \rightarrow R,-1$
 
$R,),0 \rightarrow Z,0$

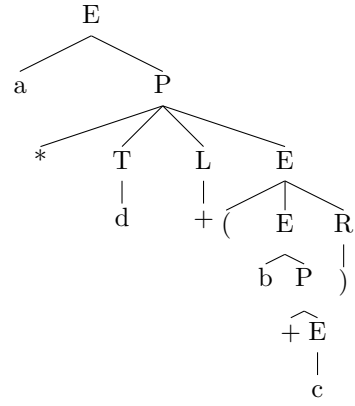
\columnbreak
\Tree[.E [.a ]
          [.P [.* ] [ .T d ] [ .L + ]
                [.E [.( ] [.E [ .b ] [ .P [ .+ ] [ .E c ] ] ] 
                [ .R ) ] ] ] ] 
\end{multicols} 
                    
A grammar defining arithmetic expressions, i.e. the same language as in our example, is used as an illustrating example in \cite{Mohri01}. Its approximating finite automaton according to \cite{Mohri01} has only two states, and thus, its acceptance transition sequences do not carry much syntactic information. 

\textbf{Definition.} \textit{Nonterminal $A$ from a CF grammar in GNF is called regular if every $E \in \mathcal{R}(A)$ has productions of the forms $E \rightarrow f$, $E \rightarrow f F$ only, or if all nonterminals from the right-hand side of every $A$ production are regular.}

This definition of regular nonterminals generalizes the definition given earlier for LID grammars. As before, regular nonterminals can be efficiently identified. If the start nonterminal is regular, then the grammar is regular. Productions $A \rightarrow b B_1 ... B_m$ where $m > 1$ are the source of inaccuracy in the approximation of CF grammars by OC automata. Following the procedure for LID grammars, we can eliminate the productions $A \rightarrow b B_1 ... B_m$ in which $B_1,...,B_{m-1}$ are regular nonterminals. 

Let us eliminate such productions $A \rightarrow b B_1 ... B_m$ that  every $E \in \mathcal{R}(B_1) \cup ... \cup \mathcal{R}(B_{m-1})$ has productions of the forms  $E \rightarrow f$, $E \rightarrow f F$ only. For $i=1,...,m-1$, we replicate all $D \in \mathcal{R}(B_i)$ along with their productions. This replication is done individually for every $i$. For every replicated nonterminal $E \in \mathcal{R}(B_i)$, we replace productions $E \rightarrow f$ with $E' \rightarrow f B_{i+1}'$ where $E'$ and $B_{i+1}'$ are the replicas of the respective nonterminals. After that, we replace $A \rightarrow b B_1 ... B_m$ with $A \rightarrow b B_1'$ where $B_1'$ is the replica of $B_1$. Clearly, this transformation does not change the language defined by the grammar. This transformation does not affect other productions $C \rightarrow d D_1 ... D_n$ with $n > 1$, and no new such productions are introduced. By applying this transformation iteratively, we can eliminate all such productions $A \rightarrow b B_1 ... B_m$ that $B_1,...,B_{m-1}$ are regular.  

The approximation of CF languages by OC automata presented earlier can be modified to approximate by finite automata. Instead of generating transitions of a OC automaton, we can generate transitions of a finite automaton. The transitions of the approximating finite automaton are obtained from the OC automaton transitions by merely stripping counter conditions and counter change functions. Proposition 4 will hold for the approximating finite automata. The proof is similar to the proof of Proposition 1. Of course, LID languages can be approximated by finite automata as well.

The approximation by finite automata will be in general less precise than the approximation by OC automata. The acceptance transition sequences of finite automata are less likely to contain matching transitions originating from the same productions. We also loose the ability to generate trees from acceptance transition sequences.

\bibliographystyle{splncs}
{\small
\bibliography{LanguageApproximation}}

\end{document}